\documentclass[12pt]{iopart}
\usepackage{iopams,amssymb,amsgen,amsfonts,amsbsy,graphicx,multirow}

\begin{document}

\bibliographystyle{unsrt}

\title[]{Nanopatterning of Si surfaces by normal incident ion erosion: influence of metal incorporation on surface morphology evolution}

\author{Jing Zhou$^{1,2}$, Stefan Facsko$^{1}$, Ming Lu$^{2}$, Wolfhard M\"{o}ller$^{1}$}
\address{$^{1}$Institute of Ion Beam Physics and Materials Research, Forschungszentrum Dresden-Rossendorf, PO Box 510119, D-01314 Dresden, Germany}
\address{$^{2}$Department of Optics Science and Engineering, Fudan University, Shanghai 200433, China}
\ead{s.facsko@fzd.de}

\begin{abstract}
The morphology evolution of Si (100) surfaces under 1200 eV Ar$^+$ ions bombardment at normal incidence with and without metal incorporation is presented. The formation of nanodot patterns is observed only when the stationary Fe concentration in the surface is above $8\times10^{14}$ cm$^{-2}$. This is interpreted in terms of an additional surface instability due to non-uniform sputtering in connection with metal enrichment at the nanodots. At low metal concentration, smoothing dominates and pattern formation is thus inhibited. The transition from a $k^{-2}$ to a $k^{-4}$ behavior in the asymptotic power spectral density function supports the conclusion that ballistic smoothing and ion-enhanced viscous flow are the two dominant mechanisms of surface relaxation.
\end{abstract}
\maketitle

\section{Introduction}
Low energy ion sputtering is a promising candidate for the fabrication of self-ordered nanostructures on both crystalline and amorphous surfaces \cite{ChasonJAP}. It was observed experimentally that ion sputtering at normal incidence or at oblique incidence can generate correlated dot or ripple patterns on metals \cite{ChasonPRBCu}, semiconductors \cite{ChasonGe, ChasonNIMBSi, GagoAPL2001} and insulators \cite{UmbachPRL}. The formation mechanism has been described theoretically by the competition between morphology induced surface instability and surface relaxation processes \cite{BradleyHarper, CuernoBarabasiPRL, MakeevNIMB, Munoz-GarciaPRB2008}.

Recently, it was reported by Ozaydin \etal \cite{OzaydinAPLMo} that molybdenum incorporation triggers the formation of ordered dot arrays on Si surfaces under the bombardment with Ar$^+$ ions at normal incidence while no correlated structures are generated in the absence of the impurities. Furthermore, Hofs\"{a}ss \etal \cite{Hofsaess} found that co-deposition of Au, Ag and Pt surfactants generates novel patterns and nanostructures on Si surfaces during ion sputtering which are absent without co-deposition. Moreover, Sanchez-Garcia \etal \cite{SanchezNanotech} observed a transition from hole to dot patterns on Si by tuning the amount of metal incorporation during ion sputtering at normal incidence. Additionally, Shenoy \etal \cite{ShenoyPRL} proposed that preferential sputtering can generate modulation both in surface morphology and composition. These findings add a new aspect to the present understanding of the development of surface topography during ion sputtering and suggest more detailed studies which address the involved roughening and smoothing mechanisms. Especially quantitative data on the influence of the metal incorporation on the morphology evolution are still missing.

In this contribution, a comparative experiment is described, which proves the important role of metal incorporation in triggering dot pattern formation on Si(100) surfaces and allows us to determine the Fe concentration at which the morphology evolution changes from roughening to smoothing.. Furthermore, the morphology evolution in the absence of metal incorporation clearly indicates that the relaxation process includes both $k^2$- and $k^4$-dependent smoothing mechanisms, with $k$ being the surface wavenumber.

\section{Experiment}
\label{sec:experiment}

The experiments were performed in an ultra-high-vacuum chamber with a base pressure of $1\times10^{-8}$ mbar. For generation of a low-pressure plasma, an inductively coupled RF plasma (ICP) generator is attached using a planar coil behind a quartz window of 150 mm diameter, which is immersed into the chamber \cite{Chevolleau2000}. At an Ar pressure of 4$\times$10$^{-3}$ mbar, the absorbed RF power was 100 W at a frequency of 13.56 MHz. Ions were extracted from the plasma by applying a 1200 eV negative DC bias to a flat sample holder which is mounted opposite to the quartz window at a distance of 10 cm. The electric field geometry and the low pressure guarantee an almost ideal normal incidence of the ions \cite{lieberman2005}. The resulting current density was 200 $\mu$A cm$^{-2}$ at the sample. During operation, the sample temperature remained below 100 $^\circ$C. It is well established that the ion induced pattern formation on Si does not depend significantly on temperature between $\sim$20 $^\circ$C and $\sim$150 $^\circ$C \cite{SanchezNanotech}.

Commercially available epi-polished Si(100) wafers (n-type, $\sim$10 $\Omega\cdot$cm) were used as samples. The metal incorporation was switched on and off by using two different masks that also served to fix the samples on the sample holder. Metal atoms sputtered from a stainless steel mask may be re-deposited on the sample surface due to partial ionization in the plasma. Alternatively, the steel mask was covered by an evaporated layer of Si of $\sim$1 $\mu$m thickness, so that the metal incorporation is largely suppressed. For post-sputtering experiments, the Si(100) surfaces were pre-patterned by ion bombardment with the same parameters as stated above, removed from the chamber and re-mounted with the Si covered mask for subsequent irradiation at different fluences.

The surface topography was analyzed {\it ex situ} by atomic force microscopy (AFM) in tapping mode. The surface metal content was analyzed by Rutherford backscattering spectrometry (RBS) using a beam of 1.7 MeV He$^+$ ions at normal incidence. Ions scattered at an angle of $170^\circ$ were analyzed using a Si surface barrier detector.

\section{Results and discussion}
\subsection{Influence of metal incorporation on dot pattern formation}
\label{subsec:comparative study}
Different samples, which will be denoted as A and B in the following, were fixed with a bare stainless steel mask and the Si covered mask, respectively, as described above. They were irradiated during 15 minutes each, corresponding to an ion fluence of $1.1\times10^{18}$ cm$^{-2}$.

\begin{table}
\caption{\label{table1}Surface contents of Ar, Fe, and Cr as obtained from RBS measurements.}

\begin{indented}
\lineup
\item[]\begin{tabular}{@{}*{5}{cccc}}
\br
&&\centre{3}{Contents (10$^{15}$ cm$^{-2}$)}\\
\ns
&&\crule{3}\\
Sample&Mount&Ar&Fe&Cr\\
\mr
A&Stainless Steel&1.63$\pm$3\%&1.53$\pm$3\%&0.40$\pm$13\%\\
B&Si coverage&1.64$\pm$3\%&0.10$\pm$50\%&0.01$\pm$100\%\\
\br
\end{tabular}
\end{indented}
\end{table}

\begin{figure}[h]
\begin{center}
\includegraphics[width=0.65\linewidth]{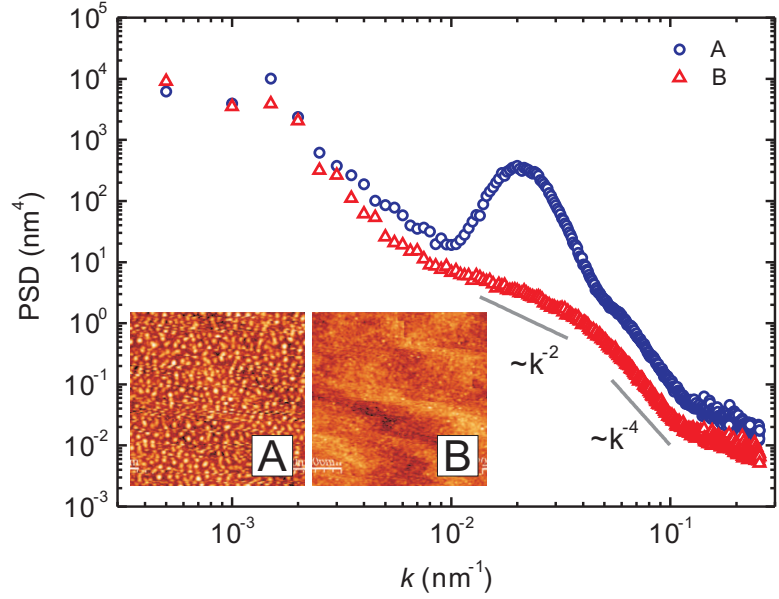}
\caption{AFM images (inset) and the corresponding PSD functions from Si(100) samples after sputtering in the presence and absence of metal incorporation (A and B, respectively). Two lines with slopes -2 and -4 are plotted for comparison (see text).}
\label{fig:comparison}
\end{center}
\end{figure}

Table \ref{table1} shows the contents of three representative elements on both samples after irradiation. The Ar content is not influenced by the amount of the metal impurities while the amount of Fe and Cr is reduced on sample B by more than one order of magnitude. The relative amounts of Fe and Cr correspond to the atomic fraction of Cr in SS 304 ($\sim$18 \%). Figure \ref{fig:comparison}  shows the power spectral density (PSD) functions obtained from the AFM images (inset) of the two samples. An ordered nanodot array is formed only in the condition of high metal incorporation, with a mean roughness of 0.85 nm. The presence of the peak in the PSD function reveals a characteristic periodicity of 50 nm. In contrast, no correlated structures are observed on sample B at a roughness of 0.23 nm.

Basically, the surface morphology evolution results from an interplay between surface instabilities and relaxation processes. According to Bradley and Harper (BH) \cite{BradleyHarper}, the morphology evolution can be described by a partial differential equation of the surface height $h(\mathbf{x},t)$ where $\mathbf{x}$ denotes the two-dimensional lateral surface coordinate and $t$ the time. At normal incidence, the laterally isotropic BH equation reads,
\begin{equation}
\label{eq:lin_eq}
\frac{\partial h(\mathbf{x},t)}{\partial t}=-v_0+\nu \nabla^2 h-D \nabla^4 h.
\end{equation}
$v_0$ is the erosion rate of the unperturbed planar surface. $\nu$ and $D$ are parameters which depend on the experimental conditions. The second term on the right hand side represents the curvature dependent roughening and smoothing for negative and positive values of $\nu$, respectively. Previous studies of low energy ion sputtering on Si surfaces suggest that the surface processes related to the second order derivative of the surface height are the curvature dependent erosion (CDE) rate \cite{BradleyHarper, SigmundJMS1973} and ballistic smoothing \cite{CarterPRBball, MoselerScience, FrostJPCMrev, VauthMayr2007}. The third term stands for a combination of relaxation processes while only ion-enhanced viscous flow (IVF) \cite{UmbachPRL} are relevant for low energy ion sputtering on Si surfaces at room temperature \cite{VauthMayr2007, ZiberiPRB}.
Equation \ref{eq:lin_eq} is linear and can be solved by Fourier Transformation, yielding for the (angular averaged) PSD function \cite{FrostJPCMrev, MakeevNIMB}
\begin{equation}
\label{eq:PSD}
\eqalign{\mathrm{P}(k,t)=\mathrm{P}(k,0)\exp(2r_kt),\\
r_k=-\nu k^2-Dk^4,}
\end{equation}
where $k$ denotes the spatial frequency, and $r_k$ the corresponding growth rate. In the case of $\nu<0$, $r_k$ has a positive maximum at $k_c=\sqrt{-\nu/2D}$, leading to a correlated pattern with the characteristic frequency $k_c$.

\begin{figure}[h]
\begin{center}
\includegraphics[width=0.6\linewidth]{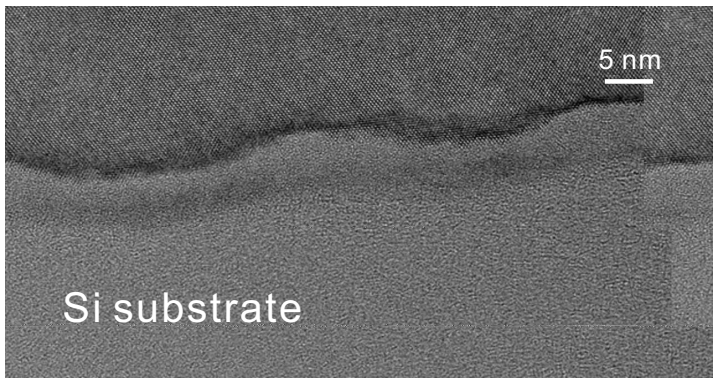}
\caption{Cross-sectional HRTEM image of the correlated nanodots generated with metal incorporation. An amorphous top layer is seen which is darkened on top of the crests due to atomic number contrast.}
\label{fig:HRTEM}
\end{center}
\end{figure}

However, in contrast to the present results, the surface processes described above cannot directly be related to the presence of impurities at the surface. Shenoy \etal \cite{ShenoyPRL} have shown that differences in sputtering yields and in surface diffusivities of different surface components can lead to modulations in topography and composition. In connection with the present experiments, a compositional inhomogeneity is indicated by the cross-sectional high-resolution transmission electron microscopy image of the dot patterned Si(100) surface shown in figure \ref{fig:HRTEM}. The nanodots are crystalline bumps with an amorphous top layer of $\sim4$ nm. The height and the periodicity of the dots are $\sim5$ nm and $\sim50$ nm, respectively. The intensity contrast is attributed to an accumulation of the metal impurities on top of the bumps. For a quantitative estimation, the darkened volume of $\sim$2.5 nm height contains $\sim$1.25$\times$10$^{16}$ Si atoms cm$^{-2}$. The areal density of metals concentrated on the crests can be estimated to be 2-3 times larger than the average total metal areal density of about 2$\times$10$^{15}$ cm$^{-2}$ (see table 1). Thus, the composition is not inconsistent with that of a stoichiometric metal disilicide. This would confirm the conjecture by Ozaydin \etal \cite{OzaydinAPLMo} that Mo silicides might be present during their ion sputtering experiments on Si surfaces. For the present system, a local decrease of the sputtering yield at the contaminated areas may be anticipated due to the larger surface binding energy of the silicide compared to pure silicon. Binary collision computer simulations using the TRIDYN program \cite{Moeller1988, Moeller2001} confirm an decrease of the Si sputtering yield from FeSi$_2$ vs. Si by $\sim$ 10 \%. Thus, once the non-uniform contamination is established, the differences of the sputter yield will indeed promote the pattern formation. There might be additional collisional effects which influence the pattern formation under metal contamination. The transition from holes to dots found by Sanchez-Garcia \etal \cite{SanchezNanotech} was qualitatively attributed to a change in the shape of collision cascade due to the metal impurities in the near-surface region.

As an alternative or additional possible mechanism, local stress might influence the dot pattern formation during ion sputtering. Previous studies show that the stress generated in the near-surface region is large enough to influence the morphology evolution \cite{ChanChasonJVSTA, ChanChasonNIMB, Kalyanasundaram2006, Dahmen2003}. Recently, Ozaydin \etal \cite{OzaydinJVSTB} found a tensile stress which they attributed to Mo incorporation during ion sputtering. Such tensile stress could cause the growth of surface protrusions as a consequence of relaxation \cite{OzaydinJVSTB}. This might even be promoted by the non-uniform contamination described above. Nevertheless, it remains speculative if a stress mechanism could also contribute to the present findings.

Thus, there is no conclusive information on the role of the metal incorporation neither from previous studies nor the present one. On the basis of the results presented above, we conclude that metal incorporation induces a new surface instability in addition to the curvature dependent erosion rate. Assuming that this new instability can be expressed as well by a second order derivative term as $\nu_{\mathrm{m}} \nabla ^2 h$, we rewrite the growth coefficient $r_k=-(\nu_{\mathrm{CDE}}+\nu_{\mathrm{m}}+\nu_{\mathrm{b}}) k^2-Dk^4$, where $\nu_{\mathrm{CDE}}$ and $\nu_{\mathrm{b}}$ represent the curvature dependent erosion rate and ballistic smoothing, respectively. Correlated patterns are generated when $\nu_{\mathrm{CDE}}+\nu_{\mathrm{m}}+\nu_{\mathrm{b}}<0$ which means that the combination of the two instabilities dominate the $k^2$-dependent surface process and compete with $k^4$-dependent smoothing mechanisms. In the absence of metal incorporation no correlated pattern is formed on the surface, indicating that the instability induced by curvature dependent erosion rate is overwhelmed by the relaxation processes. It is worth to note that the PSD function of sample B shown in figure \ref{fig:comparison} exhibits a transition from a $k^{-2}$ to a $k^{-4}$ behavior at the spatial frequency of approximately 0.05 nm$^{-1}$, as indicated by the two gray lines. This observation is in agreement with the asymptotic PSD function in the case of surface smoothing ($r_k<0$ for the entire spectrum) \cite{FrostJPCMrev, MakeevNIMB}
\begin{equation}
\label{eq:asymp}
\mathrm{P}(k,t \rightarrow \infty)\propto r_k^{-1},
\end{equation}
where the PSD function is approximately proportional to $k^{-2}$ at small frequencies and $k^{-4}$ at large frequencies.

\subsection{Smoothing of the pre-patterned surface}
For further investigation of the smoothing mechanisms, patterned samples were produced by ion sputtering with metal contamination as described above, and subsequently irradiated without metal contamination.

\begin{figure}[h]
\begin{center}
\includegraphics[width=0.65\linewidth]{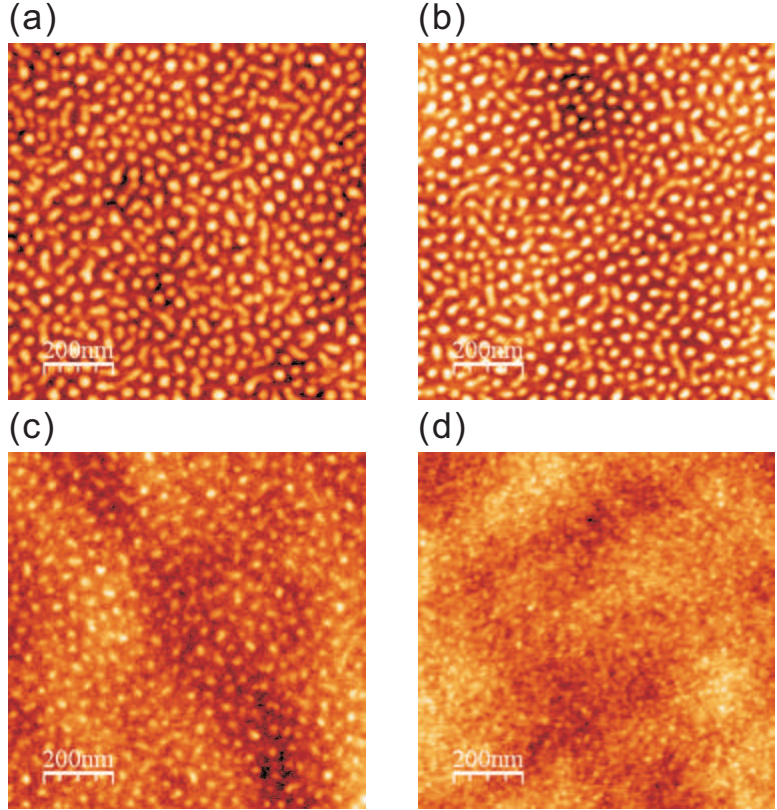}
\caption{1$\mu$m$\times$1$\mu$m AFM images of Si(100) surfaces pre-patterned by 1200-eV Ar$^+$ ions at normal incidence with a stainless steel mask (a) and post sputtered with a Si covered mask at fluences of $1.0\times10^{17}$ cm$^{-2}$ (b), $2.0\times10^{17}$ cm$^{-2}$ (c) and $3.0\times10^{17}$ cm$^{-2}$ (d).}
\label{fig:dotpattern3}
\end{center}
\vspace{0cm}
\end{figure}

\begin{figure}[h]
\begin{center}
\includegraphics[width=0.65\linewidth]{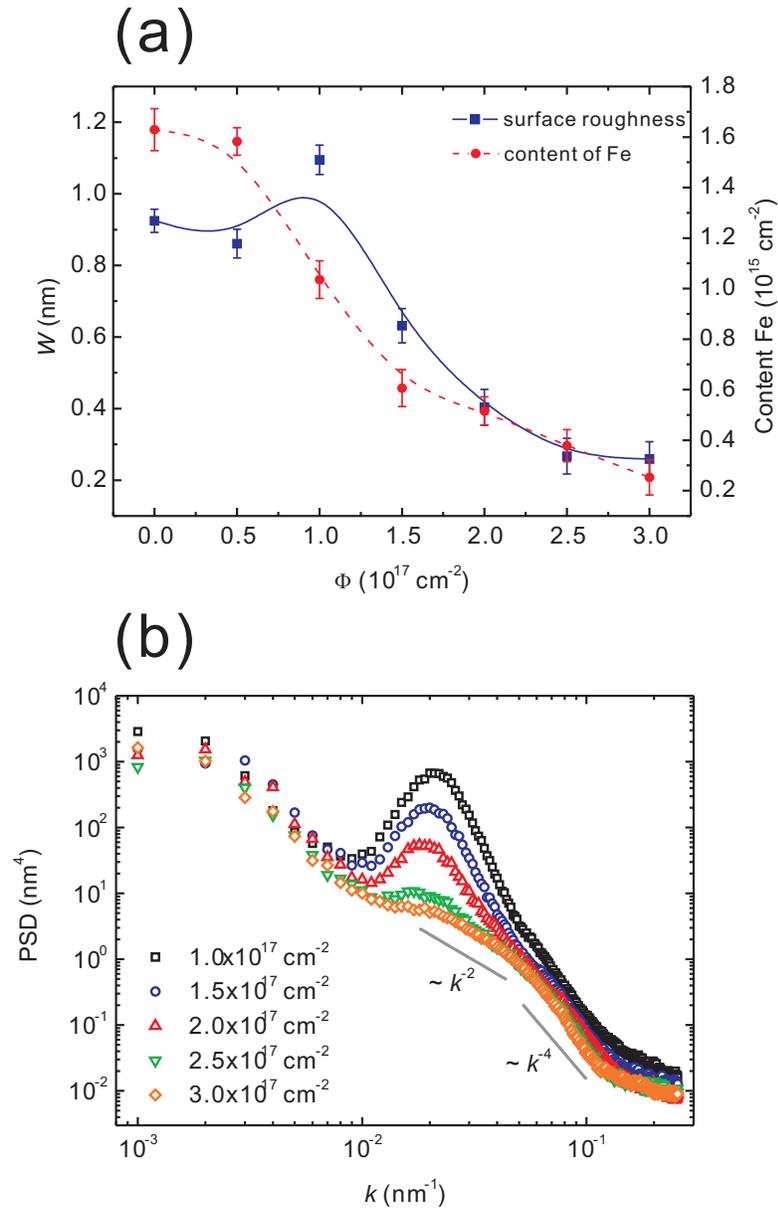}
\caption{Post-sputtering of Si(100) surfaces pre-patterned by ion sputtering under metal incorporation. (a) Root mean square (rms) surface roughness $W$ (squares) and surface content of Fe (circles) versus fluence $\Phi$. The solid and dashed lines are plotted to guide the eyes. (b) PSD functions (calculated from 1 $\mu$m$\times$1 $\mu$m AFM measurement) of the samples post-sputtered with the indicated fluences.}
\label{fig:metalpreroughen}
\end{center}
\vspace{0cm}
\end{figure}

Figure \ref{fig:dotpattern3} show the AFM images from the original surface (a), which is similar to inset A in Fig. 1, and after post-sputtering with three different fluences of $1.0\times10^{17}$ cm$^{-2}$, $2.0\times10^{17}$ cm$^{-2}$ and $3.0\times10^{17}$ cm$^{-2}$ ((b)-(d)), respectively. The evolution of the root mean square (rms) surface roughness obtained from 1 $\times$ 1 $\mu$m$^2$ AFM images and of the Fe surface content obtained from RBS are shown in figure \ref{fig:metalpreroughen} (a). The morphology evolution can be divided into two regimes: At fluences smaller than $1.0\times10^{17}$ cm$^{-2}$, the surface morphology does not visibly change, as shown by figure \ref{fig:dotpattern3} (a) and (b). The roughness remains almost constant and even slightly increases. At fluences larger than $1.0\times10^{17}$ cm$^{-2}$, the nanodot pattern gradually decays with the inter-dot distance being unchanged until the surface becomes flat. An exponential decay of the roughness is observed. The smoothing effect in this regime is also confirmed by figure \ref{fig:dotpattern3} (b)-(d).

The non-monotonic behavior in morphology evolution is attributed to the surface metal content. Based on the observation that the surface metal content decreases with ion fluence (see Fig. 4 (a) for the Fe surface content), we assume that at the beginning of post-sputtering the metal atoms inherited from the pre-patterning process still remain on the sample surface although further metal incorporation is now prevented. Therefore, the surface instability is still large enough to compete with the smoothing mechanisms. Since the surface morphology of the pre-patterned samples has reached the saturation regime, the pattern does not change in the early stage of post-sputtering at fluences which are small compared to the pre-irradiation fluence. With post-sputtering being continued, the incorporated metal atoms are gradually removed from the surface, so that finally the smoothing mechanisms dominate the morphology evolution. In this sense, a threshold value of the metal content is expected for the roughening/smoothing transition. When the metal content is larger than this threshold value, the surface instability is large enough to trigger dot pattern formation. The present experiments indicate that this threshold value for a metal impurity of mass $\sim$ 55 amu in Si is between $0.6\times10^{15}$ cm$^{-2}$ and $1\times10^{15}$ cm$^{-2}$, for Ar ion irradiation under the present conditions.

Figure \ref{fig:metalpreroughen} (b) shows the evolution of the PSD function in the decay regime. At the first sight, we find smoothing over the entire spectrum with a rapid decrease of high frequencies and a successive decrease of lower frequencies. Finally, the morphology reaches a steady state. As expected, the steady-state PSD function shows a similar behavior as the one obtained after sputtering of a virgin sample without metal contamination (see Fig. 1, curve B). According to the discussion in section \ref{subsec:comparative study}, the transient from a $k^{-2}$ to a $k^{-4}$ behavior in the PSD function corresponds to the asymptotic solution (equation (\ref{eq:asymp})) in the case of surface smoothing. This is a strong indication that the relaxation processes include both $k^2$- and $k^4$-dependent smoothing mechanisms.

Surface smoothing during ion sputtering has been investigated in several studies. Vauth and Mayr \cite{VauthMayr2007} suggest that ion-enhanced viscous flow or ballistic smoothing dominates for high or low spatial frequencies, respectively, during keV ion smoothing of amorphous surfaces. Although Frost \etal \cite{FrostJPCMrev} observed only a $k^2$-dependent surface smoothing during ion sputtering on Si, they also found first evidence for a transition from a ballistic smoothing ($\propto k^2$) to viscous flow smoothing ($\propto k^4$) in the PSD function composed from two AFM images with different scanning size. IVF has been suggested in several studies as the dominant $k^4$-dependent smoothing mechanism during low-energy ion sputtering on Si surfaces \cite{VauthMayr2007, ZiberiPRB, AlkemadeAPL}. The concept of ballistic smoothing was firstly put forward by Carter and Vishnyakov \cite{CarterPRBball} as a result of the net displacement of the forward recoils moving preferentially along the ion beam direction. The ion-impact-induced downhill current mechanism proposed by Moseler \etal \cite{MoselerScience} on the base of molecular dynamics computer simulation is considered phenomenologically virtually indistinguishable from Carter and Vishnyakov's proposal \cite{DavidovitchPRB}. Zhou \etal \cite{HZhou2008} observed $k^2$-dependent smoothing on Al$_2$O$_3$ surfaces during Ar$^+$ ion erosion at normal incidence. They attribute this effect to the ballistic atomic downhill current \cite{MoselerScience}. Accordingly and based on the discussion above, we suggest that ballistic smoothing and IVF are the two dominant relaxation processes during normal incidence Ar$^+$ ion bombardment of Si surfaces.

\section{Conclusions}
In conclusion, with 1200-eV Ar$^+$ ions bombarding Si(100) surfaces at normal incidence and below 100 $^\circ$C, we have observed dot pattern formation and surface smoothing in the presence and absence of metal incorporation, respectively. Metal incorporation corresponding to less than one monolayer critically influences the roughening/smoothing behavior. In addition to the well-known curvature-dependent erosion rate, an additional surface instability is generated, which is consistent with laterally non-uniform sputtering due to preferential accumulation of metal atoms at the nanodots. In the case of smoothing, a transition from a $k^{-2}$ to a $k^{-4}$ behavior is observed in the asymptotic PSD function, which is ascribed to ballistic smoothing and ion-enhanced viscous flow as the dominant surface relaxation mechanisms.

\ack
The authors acknowledge the experimental assistance of Dr. R. Gr\"{o}tzschel, Dr. A. Rogozin, Dr. B. Schmidt, Dr. A. M\"{u}cklich and D. Hanf. This work has been supported by DFG FOR 845 and by NSFC through Grant No. 60638010 and No. 60776038. The life cost of Jing Zhou's research stay in Dresden is supported by China Scholarship Council.

\section*{References}
\bibliography{ref}

\end{document}